\documentclass[journal]{IEEEtran}

\usepackage{graphicx} 
\usepackage{multirow}
\usepackage{subfigure}
\usepackage{color}
\usepackage{hhline}
\usepackage{xcolor}
\usepackage{transparent}
\usepackage{url}

\begin{document}

\title{Speech Enhancement based on Denoising Autoencoder with Multi-branched Encoders}

\author{Cheng Yu{*}, Ryandhimas E. Zezario{*}, Syu-Siang Wang, Jonathan Sherman, Yi-Yen Hsieh, \\ Xugang Lu, Hsin-Min Wang, ~\IEEEmembership{Senior Member,~IEEE, } and Yu Tsao, ~\IEEEmembership{Senior Member,~IEEE }
 
\thanks{Cheng Yu, Ryandhimas E. Zezario, Syu-Siang Wang, and Yu Tsao are with Research Center for Information Technology Innovation, Academia Sinica, Taipei, Taiwan, corresponding e-mail: (yu.tsao@sinica.edu.tw). }
\thanks{Jonathan Sherman is with Taiwan International Graduate Program (TIGP), Academia Sinica, Taipei, Taiwan.}
\thanks{Yi-Yen Hsieh is with Graduate Institute of Electronics Engineering (GIEE), National Taiwan University, Taipei, Taiwan.}
\thanks{Xugang Lu is with the National Institute of Information and Communications Technology, Tokyo, Japan.}
\thanks{Hsin-Min Wang is with Institute of Information Science, Academia Sinica, Taipei, Taiwan.}
\thanks{* The first two authors contributed equally to this work }
\thanks{}}

\markboth{}%
{Shell \MakeLowercase{\textit{et al.}}: Bare Demo of IEEEtran.cls for IEEE Journals}

\maketitle

\begin{abstract}
Deep learning-based models have greatly advanced the performance of speech enhancement (SE) systems. However, two problems remain unsolved, which are closely related to model generalizability to noisy conditions: (1) mismatched noisy condition during testing, i.e., the performance is generally sub-optimal when models are tested with unseen noise types that are not involved in the training data; (2) local focus on specific noisy conditions, i.e., models trained using multiple types of noises cannot optimally remove a specific noise type even though the noise type has been involved in the training data. These problems are common in real applications. In this paper, we propose a novel denoising autoencoder with a multi-branched encoder (termed DAEME) model to deal with these two problems. In the DAEME model, two stages are involved: training and testing. In the training stage, we build multiple component models to form a multi-branched encoder based on a decision tree (DSDT). The DSDT is built based on prior knowledge of speech and noisy conditions (the speaker, environment, and signal factors are considered in this paper), where each component of the multi-branched encoder performs a particular mapping from noisy to clean speech along the branch in the DSDT. Finally, a decoder is trained on top of the multi-branched encoder. In the testing stage, noisy speech is first processed by each component model. The multiple outputs from these models are then integrated into the decoder to determine the final enhanced speech. Experimental results show that DAEME is superior to several baseline models in terms of objective evaluation metrics, automatic speech recognition results, and quality in subjective human listening tests.
\end{abstract}


%
\IEEEpeerreviewmaketitle

\section{Introduction}
\IEEEPARstart
SPEECH enhancement (SE) aims to improve the quality and intelligibility of distorted speech signals, which may be caused by background noises, interference and recording devices. SE approaches are commonly used as pre-processing in various audio-related applications, such as speech communication \cite{STOI1}, automatic speech recognition (ASR) \cite{ASR1,wang2016joint, donahue2018exploring, ochiai2017multichannel}, speaker recognition \cite{SPer1, SPer2}, hearing aids \cite{hearingaid1, zhao2018deep, puder2009hearing}, and cochlear implants \cite{laizou1999signal, loizou2006speech, Cochlear1}. Traditional SE algorithms design the denoising model based on statistical properties of speech and noise signals. One class of SE algorithms computes a filter to generate clean speech by reducing noise components from the noisy speech signals. Essential approaches include spectral subtraction \cite{spectralsubtraction1}, Wiener filtering \cite{priori1}, and minimum mean square error (MMSE) \cite{priori2, ephraim1985speech}. 
Another class of SE algorithms adopts a subspace structure to separate the noisy speech into noise and clean speech subspaces, and the clean speech is restored based on the information in the clean speech subspace. Well-known approaches belonging to this category include singular value decomposition (SVD), generalized subspace approach with pre-whitening \cite{prewhitening}, Karhunen-Loeve transform (KLT) \cite{KLT}, and principal component analysis (PCA) \cite{PCA}. Despite being able to yield satisfactory performance under stationary noise conditions, the performance of these approaches is generally limited under non-stationary noise conditions. A major reason is that traditional signal processing-based solutions cannot accurately estimate noise components, consequently causing musical noises and suffering significant losses in both quality and intelligibility of enhanced speech. 
\par
Recent work has seen the emergence of machine learning and deep learning-based SE methods. Different from traditional methods, machine learning-based SE methods prepare a model based on training data in a data-driven manner without imposing strong statistical constraints. The prepared model is used to transform noisy speech signals to clean speech signals. Well-known machine learning-based models include non-negative matrix factorization \cite{NonMat, wilson2008speech, mohammadiha2013supervised}, compressive sensing \cite{CSense}, sparse coding \cite{SCod}, \cite{SparseCoding}, and robust principal component analysis (RPCA) \cite{RPCA}. Deep learning models have drawn great interest due to their outstanding nonlinear mapping capabilities. Based on the training targets, deep learning-based SE models can be divided into two categories: masking-based and mapping-based. The masking-based methods compute masks describing the time-frequency relationships of clean speech and noise components. Various types of masks have been derived, e.g., ideal binary mask (IBM) \cite{IBM}, ideal ratio mask (IRM) \cite{IRM}, target binary mask \cite{TBM}, spectral magnitude mask \cite{SMM} and phase sensitive mask \cite{PSM}. The mapping-based methods, on the other hand, treat the clean spectral magnitude representations as the target and aim to calculate a transformation function to map noisy speech directly to the clean speech signal. Well-known examples are the fully connected neural network \cite{SEDNN, liu2014experiments}, deep denoising auto-encoder (DDAE) \cite{DDAE,shivakumar2016perception}, convolutional neural network (CNN) \cite{FCN, FCN2, pandey2019new}, and long-short-term memory (LSTM) along with their combinations \cite{LSTM1, LSTM2}.
\par
Although deep learning-based methods can provide outstanding performance when dealing with seen noise types (the noise types involved in the training data), the denoising ability is notably reduced when unseen noise types (the noise types not involved in the training data) are encountered. In real-world applications, it is not guaranteed that an SE system always deals with seen noise types. This may limit the applicability of deep learning-based SE methods. \textcolor{black}{Moreover, a single and universal model, trained with the entire set of training data consisting of multiple conditions, may have constrained capability to perform well in specific condition, even though the condition has been involved in the training set. In SE tasks, the enhancement performance for a particular noise type (even if involved in the training data) can thus be weakened. This problem, caused by variances between different clusters of training data has been reported in many previous works \cite{bengio2009curriculum, kumar2010self, chang2017active}.} In this study, we intend to design a new framework to increase deep learning SE model generalizability, i.e., to improve the enhancement performance for both seen and unseen noise types. 
\par
Ensemble learning algorithms have proven to effectively improve the generalization capabilities of different machine learning tasks. Examples include acoustic modeling \cite{tsao2009ensemble}, image classification \cite{EMimg1}, and bio-medical analysis \cite{EMbio}. Ensemble learning algorithms have also been used for speech signal processing, e.g., speech dereverberation \cite{ensemble5} and SE. Lu et al. investigated ensemble learning using unsupervised partitioning of training data \cite{ensemble6}. Kim, on the other hand, proposed an online selection from trained models. The modular neural network (MNN) consists of two consecutive DNN modules: the expert module that learns specific information (e.g. SNRs, noise types, genders) and the arbitrator module that selects the best expert module in the online phase \cite{ModelSelectionDDAE}. In \cite{DRMoE}, the authors investigated a mixture-of-experts algorithm for SE, where the expert models enhance specific speech types corresponding to speech phonemes. Additionally, in \cite{MO1,ensemble8}, the authors proposed a joint estimation of different targets with enhanced speech as the primary task to improve the SE performance. 
\par
Despite the aforementioned ensemble learning models having achieved notable improvements, a set of component models is generally prepared for a specific data set and may not be suitable when applied to different large or small data sets with different characteristics. Therefore, it is desirable to build a system that can dynamically determine the complexity of the ensemble structure using a common prior knowledge of speech and noise characteristics or attributes. In this study, we propose a novel denoising autoencoder with multi-branched encoder (DAEME) model for SE. A dynamically-sized decision tree (DSDT) is used to guide the DAEME model, thereby improving the generalizability of the model to various speech and noise conditions. The DSDT is built based on prior knowledge of speech and noise characteristics from a training data set. In building the DSDT, we regard the speaker gender and signal-to-noise ratio (SNR) as the utterance-level attributes and the low and high frequency components as the signal-level attributes. Based on these definitions, the training data set is partitioned into several clusters based on the degree of attributes desired. Then, each cluster is used to train a corresponding SE model (termed component model) that performs spectral mapping from noisy to clean speech. After the first phase of training, each SE model is considered to be a branch of the encoder in the DAEME model. Finally, a decoder is trained on top of the multiple component models. In the testing stage, noisy speech is first processed by each component model. The multiple outputs from these models are then integrated into the decoder to determine the final enhanced speech. \textcolor{black}{That is, we intend to prepare multiple “matched” SE components models, and thus a decoder can incorporate “local” and “matched” information from the component models to perform SE for each specific noise condition. When the amount of training data is limited or the hardware resources for online processing are constrained, we can select fewer component models based on the DSDT tree (i.e., component models corresponding to the upper nodes of the DSDT tree) to train with the decoder. Therefore, the system complexity can be determined dynamically in the training stage.}
\par
Experimental results show that DAEME is superior to conventional deep learning-based SE models not only in objective evaluations, but also in the ASR and subjective human listening tests. \textcolor{black}{We also investigated different types of decoders and tested performance using both English and Mandarin datasets.} The results indicate that DAEME has a better generalization ability to unseen noise types than other models compared in this paper. Meanwhile, extending from our previous work \cite{ensemble6}, this paper further verifies the effectiveness of the signal-level attribute tree and knowledge-based decision tree.
\par
The rest of this paper is organized as follows. We first review several related learning-based SE approaches in Section II as our comparative baseline models. Then, we elaborate the proposed DAEME algorithm in Section III. In Section IV, we describe the experimental setup, report the experimental results, and discuss our findings. Finally, we conclude our work in Section V.

\section{RELATED WORKS}

Typically, in a learning-based SE task, a set of paired training data (noisy speech and clean speech) is used to train the model. For example, given the noisy speech $S_{\textit{y}}$, the clean speech $S_{\textit{x}}$ can be denoted as $S_{\textit{x}}=g(S_{\textit{y}})$, where $g(.)$ is a mapping function. The objective is to derive the mapping function that transforms $S_{\textit{y}}$ to $S_{\textit{x}}$, which can be formulated as a regression function. Generally, a regression function can be linear or non-linear. As mentioned earlier, numerous deep learning models have been used for SE, e.g., deep fully connected neural network \cite{SEDNN}, DDAE \cite{DDAE,shivakumar2016perception}, RNN \cite{PSM}, LSTM \cite{LSTM1,LSTM2}, CNN \cite{FCN, pandey2019new}, and the combination of these models \cite{tan2019learning, zhao2018convolutional}. \textcolor{black}{In this section, we first review some of these nonlinear mapping models along with a pseudo-linear transform, which will be used as independent and combined models for baseline comparisons in the experiments. Then, we review the main concept and algorithms of ensemble learning.}

\subsection{SE using a linear regression function}
In the linear model, the predicted weights are calculated based on a linear function of the input data with respect to the target data. In  spectral-mapping based SE approaches, we first convert a speech waveform into a sequence of acoustic features (the log-power-spectrogram (LPS) was used in this study). The spectral features for the clean speech and the noisy speech are denoted as $\textbf{X}$ and $\textbf{Y}$, respectively, where $\textbf{X} = [\textbf{x}_{1}, \textbf{x}_{2}, ..., \textbf{x}_{T}]$ and $\textbf{Y} = [\textbf{y}_{1}, \textbf{y}_{2}, ..., \textbf{y}_{T}]$, with ${T}$ feature vectors. When applying linear regression to SE, we assume that the correlation of the noisy spectral features, $\textbf{Y}$, and the clean spectral features, $\textbf{X}$, can be modeled by $\textbf{X}=\textbf{W}\textbf{Y}$, where $\textbf{W}$ denotes the affine transformation. The Moore-Penrose pseudo inverse \cite{LinMod}, which can be calculated using an orthogonal projection method, is commonly involved to solve the large size matrix multiplication. Thus, we can have
\begin{equation}
     \textbf{W}=(\textbf{C} + \textbf{Y}\textbf{Y}^T) ^{-1} \textbf{Y}^T\textbf{X}, \\
\end{equation} 
where $\textbf{C}$ denotes a scalar matrix.

On the other hand, neural network-based methods aim to minimize reconstruction errors of predicted data and reference data based on a non-linear mapping function. We will briefly describe some models adopted in this study below.

\subsection{SE using non-linear regression functions: DDAE and BLSTM}

The DDAE \cite{DDAE,shivakumar2016perception} has shown impressive performance in SE. In  \cite{DDAE}, the encoder of DDAE first converts the noisy spectral features to latent representations, and then the decoder transforms the representations to fit the spectral features of the target clean speech. For example, given the noisy spectral features $\textbf{Y} = [\textbf{y}_{1}, \textbf{y}_{2}, ..., \textbf{y}_{T}]$, we have:
\begin{equation} \\
    \begin{array}{l}
    q^{1}(\textbf{y}_{t}) = \sigma (\textbf{W}^{1}\textbf{y}_{t}+\textbf{b}^{1})  
    \\
   q^{2}(\textbf{y}_{t}) = \sigma (\textbf{W}^{2}q^{1}(\textbf{y}_{t})+\textbf{b}^{2}) \\
    \cdot \cdot \cdot \\
    q^{L}(\textbf{y}_{t}) = \sigma (\textbf{W}^{L}q^{L-1}(\textbf{y}_{t})+\textbf{b}^{L}) \\
    \\
    \hat{\textbf{x}}_{t} = \textbf{W}^{L+1}q^{L}(\textbf{y}_{t})+\textbf{b}^{L+1},
    \end{array} 
\end{equation} 
where $\sigma(.)$ is a non-linear activation function, $\textbf{W}^{l}$ and $\textbf{b}^{l}$ represent the weight matrix and bias vector for the $l$-th layer. From $q^{1}$(.) to $q^{L}$(.), the encoding-decoding process forms a non-linear regression function. The model parameters are estimated by minimizing the difference between the enhanced spectral features, $\hat{\textbf{X}} = [\hat{\textbf{x}}_{1}, \hat{\textbf{x}}_{2}, ..., \hat{\textbf{x}}_{T}]$, and the clean spectral features, $\textbf{X} = [\textbf{x}_{1}, \textbf{x}_{2}, ..., \textbf{x}_{T}]$. As reported in \cite{ensemble5}, the DDAE with a highway structure, termed HDDAE, is more robust than the conventional DDAE, and thus we will focus on the HDDAE model in this study. \textcolor{black}{This HDDAE model includes a link that copies the front hidden layers to the later hidden layers to incorporate low-level information into the supervised stage.}
\par
BLSTM models \cite{LSTM1, LSTM2} provide bilateral information exchange between series of parallel neurons, proving effective for dealing with temporal signals. The output activation from the previous layer $\textbf{h}^{l-1}_{t}$ and the activation of the previous time frame $\textbf{h}^{l}_{t-1}$ are concatenated as the input vector $\textbf{\textit{m}}^{l}_{t} = [(\textbf{h}^{l-1}_{t})^{T},(\textbf{h}^{l}_{t-1})^{T}]^{T}$ for the $l$-th layer at time frame $t$. The equations within a memory block, according to \cite{chollet2015keras}, can then be derived as follows:  
\begin{equation}
\begin{array}{lcl} \\
   \textrm{forget gate:} \ \ \textbf{\textit{f}}^{l}_{t} = \sigma_g (\textbf{W}^{l}_{f}\textbf{\textit{m}}^{l}_{t}+\textbf{U}^{l}_{f}\textbf{\textit{c}}^{l}_{t-1}+\textbf{b}^{l}_{f}) \\
   \textrm{input gate:} \ \ \textbf{\textit{i}}^{l}_{t} =\sigma_g (\textbf{W}^{l}_{i}\textbf{\textit{m}}^{l}_{t}+\textbf{U}^{l}_{i}\textbf{\textit{c}}^{l}_{t-1}+\textbf{b}^{l}_{i}) \\
   \textrm{cell vector:} \ \ \textbf{\textit{c}}^{l}_{t} = \textbf{\textit{f}}^{l}_{t} \circ \textbf{\textit{c}}^{l}_{t-1}+\textbf{\textit{i}}^{l}_{t} \circ \sigma_c(\textbf{W}^{l}_{c}\textbf{\textit{m}}^{l}_{t}+\textbf{b}^{l}_{c}) \\
   \textrm{output gate:} \ \ \textbf{\textit{o}}^{l}_{t} =\sigma_g (\textbf{W}^{l}_{o}\textbf{\textit{m}}^{l}_{t}+\textbf{U}^{l}_{o}\textbf{\textit{c}}^{l}_{t}+\textbf{b}^{l}_{o}) \\
   \textrm{output activation:} \ \ \textbf{h}^{l}_{t} = \textbf{\textit{o}}^{l}_{t} \circ \sigma_h(\textbf{\textit{c}}^{l}_{t}),    
\end{array} 
\end{equation}
where $\sigma_g$, $\sigma_c$, and $\sigma_h$ are activation functions; $\circ$ denotes the Hadamard product (element-wise product).
Finally, an affine transformation is applied to the final activation $\overrightarrow{\textbf{h}_{t}^{L}}$ and $\overleftarrow{\textbf{h}_{t}^{L}}$ of the $L$-th layer in both directions on the time axis as: 
\begin{equation}
\hat{\textbf{x}}_{t} = \overrightarrow{\textbf{W}^{L+1}}\overrightarrow{\textbf{h}_{t}^{L}} + \overleftarrow{\textbf{W}^{L+1}}\overleftarrow{\textbf{h}_{t}^{L}} + \textbf{b}^{L+1},
\end{equation}
where \overrightarrow{\textbf{W}^{L+1}} and \overleftarrow{\textbf{W}^{L+1}} are transformation matrices. 
\subsection{Ensemble learning}
Ensemble learning algorithms learn multiple component models along with a fusion model that combines the complementary information from the component models. Ensemble learning has been confirmed as an effective machine learning algorithm in various regression and classification tasks \cite{ensemble2}. In the speech signal processing field, various ensemble learning algorithms have been derived. These algorithms can be roughly divided into three categories. For the first category, the whole training set is partitioned into subsets, and each of component models is trained by a subset of the training data. Notable examples include \cite{tsao2009ensemble, ensemble6, ModelSelectionDDAE, Segmentaleigenvoice}. The second category of approaches build multiple models based on different types of acoustic features. Well-known approaches include \cite{MO1, wang2018multiobjective}. The third category constructs multiple models using different model types or structures. Successful approaches belonging to this category include \cite{ensemble8, le2013ensemble}. By exploiting complementary information from multiple models, ensemble learning approaches can yield more robust performance compared to conventional machine learning algorithms. 

\graphicspath{ {./images/} }
\begin{figure}[!t]
{\transparent{0.8}\includegraphics[width=\columnwidth]{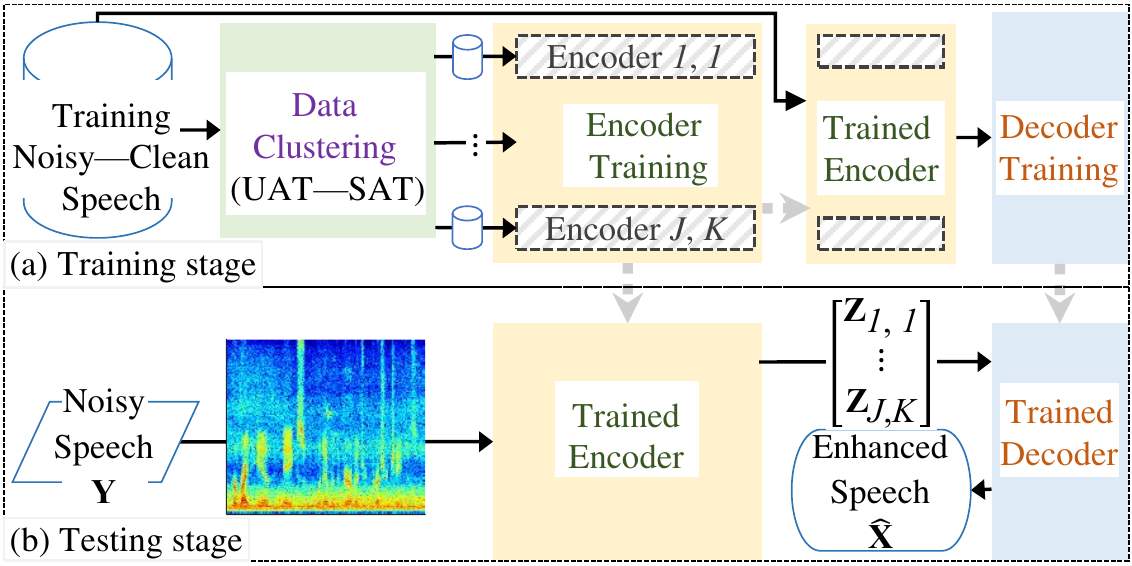}}\caption{The overall architecture of the DAEME approach.}
\label{fig:DAEME}
\end{figure}

\par
Despite the impressive performance in speech signal processing tasks, there are  three  possible  limitations to implementing a typical ensemble learning model in a real-world SE application: First, it is difficult to tell the contribution of each component model. Second, the amount of training data for training multiple models may not be sufficient since the training data for each component model is a subset of all training data. Third, it is not easy to  dynamically determine which component models should be kept and which should be removed.
\par
To overcome the above three limitations, we propose a novel DAEME SE algorithm, which comprises training and testing stages. In the training stage, a DSDT is constructed based on the attributes of speech and noise acoustic features. Since the DSDT is constructed in a top-down manner, a node in a higher layer consists of speech data with broad attribute information. On the other hand, a node in a lower layer denotes the speech data with more specific attribute information. An SE model is built for each node in the tree. \textcolor{black}{The models corresponding to higher layers of the DSDT tree are trained with more training data and can be used as initial models to estimate component models corresponding to lower layers of the DSDT tree.} These SE models are then used as component models in the multi-branched encoder. Next, a decoder is trained in order to combine the results of the multiple models. In the testing stage, a testing utterance is first processed by the component models; the outputs of these models are then combined by the decoder to generate the final output. Since the DSDT has a clear physical structure, it becomes easy to analyze the SE property of each component model. Moreover, based on the DSDT, we may dynamically determine the number of SE models according to the amount of training data and the hardware limitation in the training stage. Last but not least, in \cite{overen}, the authors proposed a special ``overfitting" strategy to interpret the effectiveness of a component model. \textcolor{black}{When training ensemble models, we intend to implement a “conditional overfitting” strategy, which aims to train each component model to overfit to (or perfectly match) its training data (i.e., the training data corresponding to each node in the tree in our work).} The DAEME algorithm has better interpretability by using attribute-fit regressions based on the DSDT.

\section{Proposed DAEME Algorithm}

\par
The overall architecture of the proposed DAEME approach is depicted in Fig. \ref{fig:DAEME}. In this section, we will detail the training and testing stages. 
\subsection{Training stage} 
\begin{figure}[ht]
\centering
\includegraphics[width=7cm]{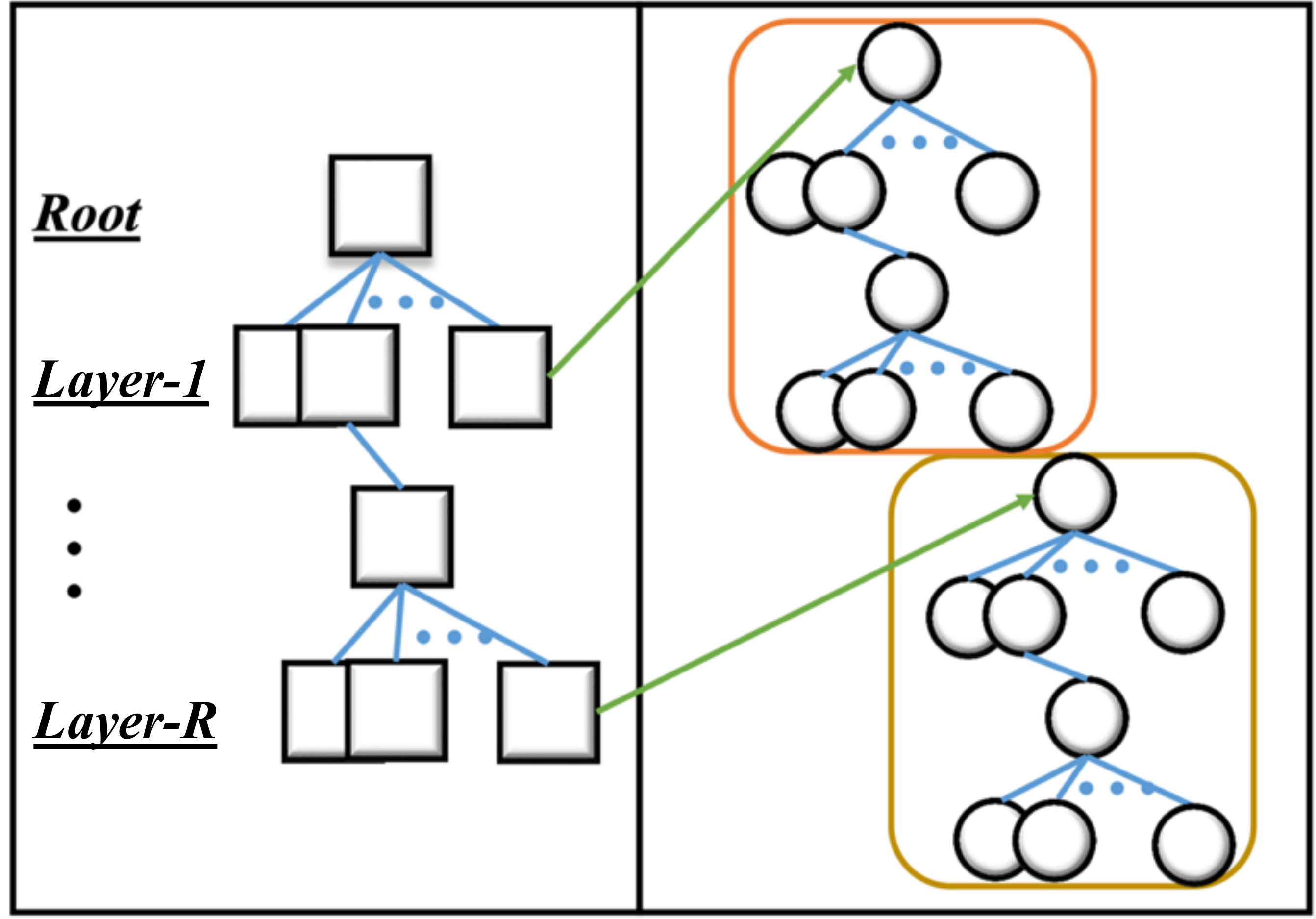} 
\caption{The tree built on the utterance-level attribute (UAT) and signal-level attribute (SAT).} 
\label{fig:USAT}
\end{figure}

In the training stage, we first build a tree based on the attributes of the training utterances in a top-down manner. The root of the tree includes the entire set of training data. Then, based on the utterance-level attributes, we create the branches from the root node. As the layers increase, the nodes represent more specific utterance-level attribute information. Next, we process signal-level attributes to create branches upon the nodes. Finally, we have a tree with multiple branches. As shown in Fig. \ref{fig:USAT}, based on the utterance-level and signal-level attributes, we build UAT and SAT, respectively. In the following subsection, we will introduce the UAT and SAT in more detail. 
\graphicspath{ {./images/} }

\subsubsection{Utterance-level attribute tree (UAT)}
The utterance-level attributes include speaker and speaking environment factors, such as the gender, age, accent, or identity for the speaker factor, and the signal-to-noise ratio (SNR), noise type, and acoustic environment for the environment factor. As reported in \cite{attribute}, three major factors that affect the SE performance are the noise type, speaker, and SNR. In real-world scenarios, the noise type is usually inaccessible beforehand. Accordingly, we only consider the speaker and SNR factors when building the attribute tree in this study. The root node includes all the training utterances. Next, we create two branches from the root node for male and female speakers. Therefore, each node in the second layer contains the training utterances of male or female speakers. Then, the two nodes in the second layer are further divided to four nodes in the third layer, each containing the training data of ``male with high SNR”, ``male with low SNR”, ``female with high SNR” or ``female with low SNR”.
\subsubsection{Signal-level attribute tree (SAT)} 
For the signal-level attributes, we segment the acoustic features into several groups, each with similar properties. For SE, it has been reported that considering low and high frequency properties can give improved performance \cite{DWPT}. In this study, we segment the acoustic features into high-frequency and low-frequency parts by two methods: spectral segmentation (SS) and wavelet decomposition (WD). With the signal-level attributes, we can create an additional layer for each node of the tree constructed using the utterance-level attributes, forming final binary branches from each node.
\subsubsection{Models of multi-branched encoder and decoder}
The overall procedure of the training stage is shown in Fig. \ref{fig:DAEME} (a). We estimate a component model for each node in the tree. The input and output of the model is paired by noisy and clean features, and the objective is to estimate the mapping function between the paired features. We reason that the model at the highest layer (the root node) characterizes the mapping function globally, rather than considering specific local information. In our case, the local information includes the speaker attributes and environment features segmented by each layer and node in the DSDT. On the other hand, the model corresponding to a node in a lower layer characterizes a more localized mapping of paired noisy and clean features. More specifically, each model characterizes a particular mapping in the overall acoustic space. Given the component SE models, we then estimate a decoder. Note that by using the component models to build multiple SE mapping functions, we can factorize the global mapping function (the mapping of the root node in our system) into several local mapping functions; the decoder uses the complementary information provided by the local mappings to obtain improved performance as compared to the global mapping. This also follows the main concept of ensemble learning: building multiple over-trained models, each specializing in a particular task, and then computing a fusion model to combine the complementary information from the multiple models. Assume that we have $J \times K$ component models built by UAT and SAT, the SE can be derived with the component models as
\begin{equation} \\
    \begin{array}{c}
    \textbf{Z}_{1,1} \ = \ F_{1,1}^{(E)} \ (\textbf{Y}) 
    \\
    \textbf{Z}_{1,2} \ = \ F_{1,2}^{(E)} \ (\textbf{Y})
    \\
    .
    .
    \\
    \textbf{Z}_{j,k} \ = \ F_{j,k}^{(E)} \ (\textbf{Y})
    \\
    .
    .
    \\
    \textbf{Z}_{J,K} \ = \ F_{J,K}^{(E)} \ (\textbf{Y})
    \end{array} 
\end{equation} 
and the decoder as
\begin{equation} \\
    \begin{array}{c}
    \hat{\textbf{X}} \ = \ F_{\theta}^{(D)}\ (\textbf{Z}_{1,1}, \textbf{Z}_{1,2}, ..., \textbf{Z}_{J,K}),
    \end{array} 
\end{equation} 
where  $\textit{j}$ and $\textit{k}$ represents the UAT node index and the SAT node index, respectively. $F_{j,k}^{(E)}$ denotes the transformation of the $j,k$-th component model, $\textbf{Z}_{j,k}$ is the corresponding output, and $F_{\theta}^{(D)}$ denotes the transformation of the decoder. The parameter, $\theta$, is estimated by:

\begin{equation}
\hat{\theta}=\mathop{\arg\min}_{\theta}D_{iff}(\mathbf{X}, \hat{\mathbf{X}}).
\end{equation}
where $D_{iff}(\cdot)$ denotes a function that measures the difference of paired inputs. In this study, we adopted the mean-square-error (MSE) function for $D_{iff}(\cdot)$.

\subsection{Testing stage} 
The testing  stage of the DAEME algorithm is shown in Fig. \ref{fig:DAEME} (b). The input speech is first converted to acoustic features, and each component model processes these features separately. The outputs are then combined with a decoder. Finally, the enhanced acoustic features are reconstructed to form the speech waveform. \textcolor{black}{As mentioned earlier, based on the tree structure, we may dynamically determine the number of ensemble models according to the online processing hardware limitation or the amount of training data available. Note that, the system complexity is determined in the training stage.} 
\par
It is important to note that one can use different types of models and different types of acoustic features to form the component models in the multi-branched encoder. In this study, we intend to focus on comparing the effects caused by different attributes of the speech utterances, so the same neural network model architecture was used for all components in the multi-branched encoder. Additionally, we adopted different types of models to form encoders and decoders to investigate the correlation between the model types and the overall performance. Please note that the knowledge, such as gender and SNR information, is only used in the training stage for building the component models. Such \textcolor{black}{prior} knowledge is not used in the testing stage.

\section{EXPERIMENTS \& RESULTS}

We used two datasets to evaluate the proposed algorithm, namely the Wall Street Journal (WSJ) corpus \cite{CSR} and the Taiwan Mandarin version of the hearing in noise test (TMHINT) sentences \cite{TMHINT}. In this section, we present the experimental setups for the two datasets and discuss the evaluation results.
\subsection{Evaluation metrics}
To evaluate the performance of SE algorithms, we used Perceptual Evaluation of Speech Quality (PESQ) \cite{PESQ1} and Short-Time Objective Intelligibility (STOI) \cite{STOIpredict}. PESQ and STOI have been widely used as standard objective measurement metrics in many related tasks \cite{SEDNN, WeinerDDAE, PODDAE}. PESQ specifically aims to measure the speech quality of the utterances, while STOI aims to evaluate the speech intelligibility. The PESQ score ranges from -0.5 to 4.5, a higher score indicating better speech quality. The STOI score ranges from 0.0 to 1.0, a higher score indicating higher speech intelligibility.

\subsection{Experiments on the WSJ dataset}
The WSJ corpus is a large vocabulary continuous speech recognition (LVCSR) benchmark corpus, which consists of 37,416 training and 330 test clean speech utterances. The utterances were recorded at a sampling rate of 16Khz. We prepared the noisy training data by artificially contaminating the clean training utterances with 100 noise types \cite{wang2013towards} at 31 SNR levels (20dB to -10dB, with a step of 1dB). Each clean training utterance was contaminated by one randomly selected noise condition; therefore, the entire training set consisted of 37,416 noisy-clean utterance pairs. To prepare the noisy test data, four types of noises, including two stationary types (i.e., car and pink) and two non-stationary types (i.e., street and babble), were added to the clean test utterances at six SNR levels (-10 dB, -5dB, 0 dB, 5 dB, 10 dB, and 15 dB). Note that the four noise types used to prepare the test data were not involved in preparing the training data. We extracted acoustic features by applying 512-point Short-time Fourier Transform (STFT) with a Hamming window size of 32 ms and a hop size of 16 ms on training and test utterances, creating 257-point STFT log-power-spectrum (LPS) features.
For the baseline SE system, we used a BLSTM model with two bidirectional LSTM layers of 300 nodes followed by a fully connected output layer \cite{PSM}.  For DAEME, similar to the baseline system, we used two bidirectional LSTM layers with 300 nodes and one fully connected output layer to form the components in the multi-branched encoder. Since an essential objective of DAEME is to provide sufficient information to be able to generalize the SE models, multiple models are created, each learning the particular mapping function. \textcolor{black}{A CNN composed of three convolutional layers and two fully connected layers, each convolutional layer containing 64 channels and each fully connected layers containing 1024 neurons, was used to form the decoder.} For a fair comparison, all of the component and decoder models are trained using the same number of epochs. 

\subsection{Prior knowledge of speech and noise structures}
\begin{figure}[ht]
 \subfigure[]{
   \includegraphics[scale =0.78]{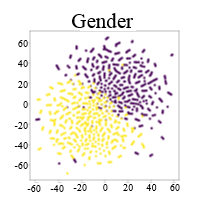}
   \label{fig:tsne-gender}
 }
 \subfigure[]{
   \includegraphics[scale =0.78]{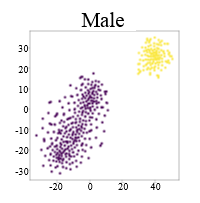}
   \label{fig:tsne-m-snr}
 }
  \subfigure[]{
   \includegraphics[scale =0.78]{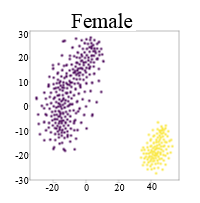}
   \label{fig:tsne-f-snr}
 }
\caption{T-SNE analysis on the utterance-level attributes. (a) Gender: male (yellow dots) vs. female (purple dots). (b) Male: high SNR (yellow dots) vs. low SNR (purple dots). (c) Female: high SNR (yellow dots) vs. low SNR (purple dots). Note that we used the LPS features to draw the T-SNE plots.}
\label{fig:TSNE}
\end{figure}

Recent monaural SE studies have mentioned the importance of attributive generalizability in SE systems. For example, in \cite{attribute}, Morten et al. compared noise types, speaker, and SNR level and their potential to enhance the intelligibility of SE systems. To build the DSDT and the corresponding DAEME system, we first qualitatively analyzed the acoustic properties of speech attributes. As mentioned earlier, we assumed that the noise type is inaccessible and conducted the T-SNE analysis \cite{tsne} on the training data of the WSJ dataset. T-SNE performs nonlinear dimensionality reduction on the original data, and the T-SNE plot has been popularly used for data representation and analysis. Like other dimensionality reduction techniques, such as PCA, the two axes may not have physical meanings. Nevertheless, we \textcolor{black}{can determine} the clusters by associating the axes with physical attributes. The analysis results are shown in Fig. \ref{fig:TSNE}, where in Fig. \ref{fig:TSNE} (a), we analyzed the gender attributes, and in Fig. \ref{fig:TSNE} (b) and (c), we analyzed the SNR attributes.   

\graphicspath{ {./images/} }

From Fig. \ref{fig:TSNE} (a), we can note  a clear separation between male (yellow dots) and female (purple dots). In Figs. \ref{fig:TSNE} (b) (i.e., male data of different SNR levels) and \ref{fig:TSNE} (c) (i.e., female data of different SNR levels), the T-SNE analysis also shows a clear separation between high SNR (10dB and above, yellow dots) and low SNR (below 10dB, purple dots) in both gender partitions.

On the other hand, from the signal point of view, most real-world noises were non-stationary in the time-frequency domain. This suggests that noise pollution is unlikely to occur in all frequency bands at the same time even under low SNR conditions. In our previous studies \cite{Segmentaleigenvoice}, segmental frequency bands were proposed to enhance speaker adaptation. In this study, the signal-level attribute is used to consider the low and high frequency bands. With the signal-level attribute, the SE algorithm can obtain more speech structure information even under noisy conditions.

\subsubsection{The effectiveness of the UAT}
We first analyzed the DAEME model with a tree built with the utterance-level attributes. As mentioned in Section III-A-1), the root node of the UAT included the entire set of 37,416 noisy-clean utterance pairs for training. The entire training set was divided into male and female in the first layer, each with 18,722 (male) and 18,694 (female) noisy-clean training pairs, respectively. For the next layer, the gender-dependent training data was further divided into low and high SNR conditions. Finally, there were four leaf nodes in the tree, each with 6,971 (female with high SNR), 11,723 (female with low SNR), 7,065 (male with high SNR), and 11,657 (male with low SNR) noisy-clean training pairs. We investigated the performance of DAEME with different numbers of components in the multi-branched encoder, and thus built three systems. The first system had two component models: male and female. The second system had four component models: female with high SNR, female with low SNR, male with high SNR, and male with low SNR. The third system had six component models: female, male, female with high SNR, female with low SNR, male with high SNR, and male with low SNR. The three systems are termed DAEME-UAT$_{(2)}$, DAEME-UAT$_{(4)}$, and DAEME-UAT$_{(6)}$, respectively.
\par
\graphicspath{ {./images/} }
\begin{figure}[t]
\centering
   {
\transparent{0.8}\includegraphics[width=8.8cm]{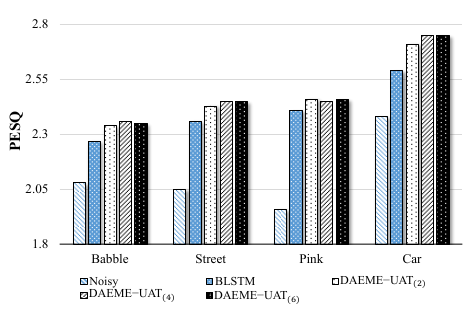}
}
    \caption{Performance comparison of DAEME-UAT$_{(i)}$, $i = 2,4,6$, and a single model BLSTM SE method in terms of the PESQ score. The results of unprocessed noisy speech, denoted as Noisy, are also listed for comparison.}
    \label{fig:UAT246PESQ}
\graphicspath{ {./images/} }
\centering
  {
\transparent{0.8}\includegraphics[width=8.8cm]{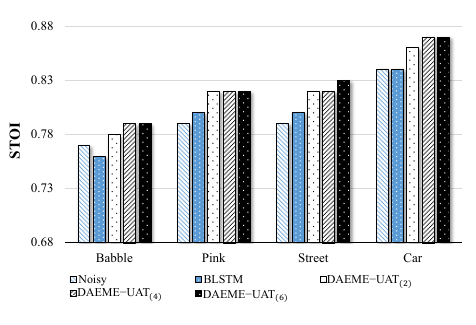}
}
    \caption{Performance comparison of DAEME-UAT$_{(i)}$, $i = 2,4,6$, and a single model BLSTM SE method in terms of the STOI score. The results of unprocessed noisy speech, denoted as Noisy, are also listed for comparison.}
    \label{fig:UAT246STOI}
\end{figure}
In the first set of experiments, we examined the effectiveness of the UAT-based DAEME systems versus single model SE algorithms. We selected a single model BLSTM for the benchmark performance. This BLSTM model was trained with the entire set of 37,416 noisy-clean utterance pairs. Fig. \ref{fig:UAT246PESQ} shows the PESQ scores of the proposed systems (DAEME-UAT$_{(2)}$, DAEME-UAT$_{(4)}$, DAEME-UAT$_{(6)}$) and the single  BLSTM model. Fig. \ref{fig:UAT246STOI} shows the STOI scores. In these two figures, each bar indicates an average score over six SNR levels for a particular noise type. From the results in Figs. 4 and 5, we note that all the DAEME-UAT models outperform the single BLSTM model methods in all noise types. More specifically, using only the UAT, the DAEME systems already surpasses the performance of the single model SE methods by a notable margin. This result justifies the tree-based data partitioning in training the DAEME method.
\par
To further verify the effectiveness of the UAT, we constructed another tree. The root node also contained the entire set of 37,416 noisy-clean utterance pairs. The training utterance pairs were randomly divided into two groups to form two nodes in the second layer, where each node contained 18,694 and 18,722 noisy-clean utterance pairs, respectively. In the third layer, the training utterance pairs of each node in the second layer were further divided into two groups randomly to form four nodes in the third layer. Since the tree was built by randomly assigning training utterances to each node, this tree is termed random tree (RT) in the following discussion. Based on this RT, we built three systems, termed DAEME-RT$_{(2)}$, DAEME-RT$_{(4)}$, and DAEME-RT$_{(6)}$, each consisting of 2, 4, and 6 component models, respectively. Figs. 6 and 7 compare the PESQ and STOI scores of different DAEME systems (including DAEME-UAT$_{(2)}$, DAEME-UAT$_{(4)}$, DAEME-UAT$_{(6)}$, DAEME-RT$_{(2)}$, DAEME-RT$_{(4)}$, and DAEME-RT$_{(6)}$) based UAT or RT. From the figures, we can draw two observations. First, DAEME-UAT$_{(6)}$ achieve a better performance than their respective counterparts with less component models. \textcolor{black}{Since the same decoder is used, the performance improvements are mainly contributed by the specific designs of component models. Based on the performance improvements, we can also see the advantage of using the tree structures to design component models in the DAEME framework.} Second, the results of DAEME-UAT$_{(i)}$ are consistently better than DAEME-RT$_{(i)}$ for $i = 2,4,6$. The result confirms the effectiveness of the UAT over RT. For ease of further comparison, we list the detailed PESQ and STOI scores of DAEME-UAT$_{(6)}$ (the best system in Figs 6 and 7) in Tables \ref{tab:PESQ_DAEME-UAT6} and \ref{tab:STOI_DAEME-UAT6}. 

\graphicspath{ {./images/} }
\begin{figure}[t]
\centering
   \includegraphics[width=8.8cm]{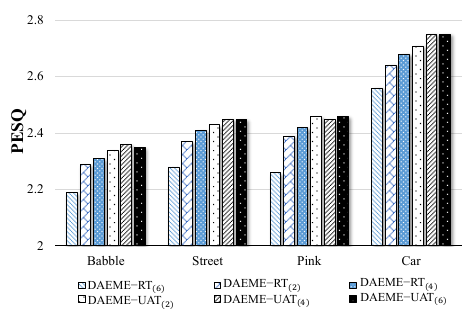}
    \caption{Performance comparison of DAEME-UAT$_{(i)}$ and DAEME-RT$_{(i)}$, $i = 2, 4, 6$ in terms of the PESQ score.}
    \label{fig:UATRTPESQ}
\end{figure}

\graphicspath{ {./images/} }
\begin{figure}[ht]
\centering
\includegraphics[width=8.8cm]{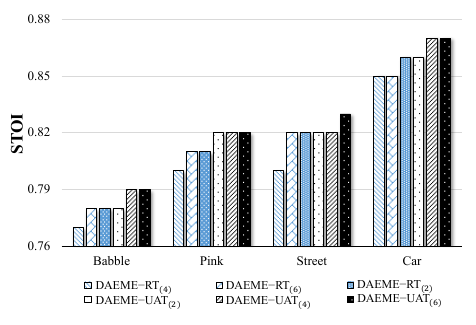}
    \caption{Performance comparison of DAEME-UAT$_{(i)}$ and DAEME-RT$_{(i)}$, $i = 2, 4, 6$ in terms of the STOI score.}
    \label{fig:UATRTSTOI}
\end{figure}

\begin{table}[b]
\caption{\scshape{P}ESQ scores of the DAEME-UAT${_6}$ system at four noise types and six SNRs. Avg. denotes the average scores.}
\begin{center} 
\hrulefill \\
\textbf{DAEME-UAT$_{(6)}$} \\
\begin{tabular}{c||c|c|c|c|c|c||c}
\hline
& \textbf{15dB} & \textbf{10dB} &  \textbf{5dB} & \textbf{0dB} & \textbf{-5dB} &  \textbf{-10dB} & \textbf{Avg.} \\
\hline
CAR & 3.21 & 3.11 & 2.96 & 2.72 & 2.41 & 2.06 & \textbf{2.75}\\
\hline
PINK & 3.16 & 2.98 & 2.73 & 2.39 & 1.96 & 1.56 & \textbf{2.46}\\
\hline
STREET & 3.15 & 2.98 & 2.73 & 2.37 & 1.92 & 1.55 & \textbf{2.45}\\
\hline
BABBLE & 3.15 & 2.93 & 2.63 & 2.20 & 1.74 & 1.43 & \textbf{2.35}\\
\hline
\hline
\textbf{Avg.} & \textbf{3.17} & \textbf{3.00} & \textbf{2.76} & \textbf{2.42} & \textbf{2.01} & \
\textbf{1.65} & \textbf{2.50}\\
\end{tabular} \\
\hrulefill
\end{center}
\label{tab:PESQ_DAEME-UAT6}

\caption{\scshape{S}TOI scores of the DAEME-UAT${_6}$ system at four noise types and six SNRs. Avg. denotes the average scores}
\begin{center} 
\hrulefill \\
\textbf{DAEME-UAT$_{(6)}$} \\
\begin{tabular}{c||c|c|c|c|c|c||c}
\hline
& \textbf{15dB} & \textbf{10dB} &  \textbf{5dB} & \textbf{0dB} & \textbf{-5dB} &  \textbf{-10dB} & \textbf{Avg.}\\
\hline
CAR & 0.93 & 0.92 & 0.90 & 0.87 & 0.82 & 0.76 & \textbf{0.87}\\
\hline
PINK & 0.94 & 0.92 & 0.90 & 0.84 & 0.74 & 0.59 & \textbf{0.82}\\
\hline
STREET & 0.94 & 0.92 & 0.89 & 0.81 & 0.68 & 0.51 & \textbf{0.79}\\
\hline
BABBLE & 0.94 & 0.92 & 0.90 & 0.84 & 0.75 & 0.62 & \textbf{0.83}\\
\hline
\hline
\textbf{Avg.} & \textbf{0.94} & \textbf{0.92} & \textbf{0.90} & \textbf{0.84} & \textbf{0.75} & \
\textbf{0.62} & \textbf{0.83}\\
\end{tabular} \\
\hrulefill
\end{center}
\label{tab:STOI_DAEME-UAT6} 
\end{table}

\subsubsection{The effectiveness of the SAT}
As introduced in Section III-A-2), we can use the SS and WD approaches to build the SAT. To compare the SS and WD approaches, we used the SAT on top of DAEME-UAT$_{(6)}$, which achieved the best performance in Fig. \ref{fig:UATRTPESQ} (PESQ) and Fig. \ref{fig:UATRTSTOI} (STOI), to build two systems: DAEME-USAT$_{(SS)(12)}$ and DAEME-USAT$_{(WD)(12)}$, where USAT denotes the system using the ``UA+SA" tree. Because the SA tree was built into each node of the UA tree, the total number of component models for an DAEME-USAT system is the number of nodes in the UA tree multiplied by the number of nodes in the SA tree. We applied the SS and WD approaches to segment the frequency bands. For SS, we directly separated the low and high frequency parts based on the spectrograms. Among the 257 dimensional spectral vector, we used [1:150] and [108:257] coefficients as the low- and high-frequency parts, respectively. For WD, we used the Biorthogonal 3.7 wavelet \cite{vetterli1995wavelets}, which was directly applied on the speech waveforms to obtain the low- and high-frequency parts. In either way, the SA tree had two nodes. Therefore, we obtained DAEME-USAT$_{(SS)(12)}$ and DAEME-USAT$_{(WD)(12)}$, respectively, based on DAEME-UAT$_{(6)}$ after applying the SA trees constructed by the SS and WD approaches. \par
The PESQ and STOI scores of DAEME-USAT$_{(SS)(12)}$ and  DAEME-USAT$_{(WD)(12)}$ are listed in Tables \ref{tab:PESQ_DAEME-UAT12} and \ref{tab:STOI_DAEME-UAT12}, respectively. From these two tables, we observe that DAEME-USAT$_{(WD)(12)}$ outperformed DAEME-USAT$_{(SS)(12)}$ in terms of the PESQ score across 15dB to -10dB SNR conditions, while the two systems achieved comparable performance in terms of the STOI score. Comparing the results in Tables I and II and the results in Tables III and IV, we note that both DAEME-USAT$_{(SS)(12)}$ and DAEME-USAT$_{(WD)(12)}$ achieved better PESQ and STOI scores than DAEME-UAT$_{(6)}$. 
In addition to the average PESQ and STOI scores, we adopted a statistical hypothesis test, the dependent t-test, to verify the significance of performance improvements \cite{Hayter}, \cite{Agresti}. We compared two methods by testing the individual average PESQ/STOI scores of 24 matched pairs (4 noise types and 6 SNR levels) and estimated the $p$-value. For the t-test, we considered $H_{0}$ as “method-II is not better than method-I,” and $H_{1}$ as “method-II is better than method-I.” Small $p$-values imply significant improvements of method-II over method-I across the 24 matched pairs. A threshold of 0.01 was used to determine whether the improvements were statistically significant. For the PESQ scores, the improvements of DAEME-UAT$_{(6)}$, DAEME-USAT$_{(SS)(12)}$ and DAEME-USAT$_{(WD)(12)}$ over BLSTM are significant with very small $p$-values. In addition, the improvements of DAEME-USAT$_{(SS)(12)}$ and DAEME-USAT$_{(WD)(12)}$ over DAEME-UAT$_{(6)}$ are significant with very small $p$-values. For the STOI scores, the improvements of DAEME-UAT$_{(6)}$, DAEME-USAT$_{(SS)(12)}$ and DAEME-USAT$_{(WD)(12)}$ over BLSTM are significant with very small $p$-values. However, the differences of DAEME-USAT$_{(SS)(12)}$ over DAEME-USAT$_{(WD)(12)}$ are not significant. The results first confirm the effectiveness of the DAEME model in providing better speech quality and intelligibility. The results also confirm the effectiveness of the SA tree in enabling DAEME to generate enhanced speech with better quality.

\begin{table}[ht]
\caption{\scshape{P}ESQ scores of the DAEME-USAT$_{(SS)(12)}$ and DAEME-USAT$_{(WD)(12)}$ systems at four noise types and six SNRs. Avg. denotes the average scores.}
\begin{center} 
\hrulefill \\
\textbf{DAEME-USAT$_{(SS)(12)}$} \\
\begin{tabular}{c||c|c|c|c|c|c||c}
\hline
& \textbf{15dB} & \textbf{10dB} &  \textbf{5dB} & \textbf{0dB} & \textbf{-5dB} &  \textbf{-10dB} & \textbf{Avg.}\\
\hline
CAR & 3.34 & 3.23 & 3.05 & 2.79 & 2.46 & 2.09 & \textbf{2.83}\\
\hline
PINK & 3.27 & 3.08 & 2.82 & 2.46 & 2.02 & \textbf{1.60} & \textbf{2.54}\\
\hline
STREET & 3.27 & 3.09 & 2.81 & 2.43 & 1.97 & \textbf{1.58} & \textbf{2.53}\\
\hline
BABBLE & 3.25 & 3.02 & 2.69 & 2.24 & 1.75 & \textbf{1.44} & \textbf{2.40}\\
\hline
\hline
\textbf{Avg.} & \textbf{3.28} & \textbf{3.11} & \textbf{2.84} & \textbf{2.48} & \textbf{2.05} & \textbf{1.68} & \textbf{2.57}
\end{tabular} \\
\hrulefill \\
\hrulefill \\
\textbf{DAEME-USAT$_{(WD)(12)}$} \\
\begin{tabular}{c||c|c|c|c|c|c||c}
\hline
& \textbf{15dB} & \textbf{10dB} &  \textbf{5dB} & \textbf{0dB} & \textbf{-5dB} &  \textbf{-10dB} & \textbf{Avg.}\\
\hline
CAR & \textbf{3.49} & \textbf{3.36} & \textbf{3.17} & \textbf{2.88} & \textbf{2.52} & \textbf{2.13} & \textbf{2.93}\\
\hline
PINK & \textbf{3.34} & \textbf{3.12} & \textbf{2.84} & \textbf{2.48} & \textbf{2.03} & 1.59 & \textbf{2.57}\\
\hline
STREET & \textbf{3.39} & \textbf{3.18} & \textbf{2.89} & \textbf{2.49} & \textbf{2.00} & \textbf{1.58} & \textbf{2.59}\\
\hline
BABBLE & \textbf{3.37} & \textbf{3.12} & \textbf{2.77} & \textbf{2.30} & \textbf{1.78} & \textbf{1.44} & \textbf{2.46}\\
\hline
\hline
\textbf{Avg.} & \textbf{3.40} & \textbf{3.20} & \textbf{2.92} & \textbf{2.54} & \textbf{2.08} & \textbf{1.69} & \textbf{2.64}

\label{tab:PESQ_DAEME-UAT12} 
\end{tabular} \\
\hrulefill
\end{center}
\caption{\scshape{S}TOI scores of the DAEME-USAT$_{(SS)(12)}$ and DAEME-USAT$_{(WD)(12)}$ systems at four noise types and six SNRs. Avg. denotes the average scores.}
\begin{center} 
\hrulefill \\
\textbf{DAEME-USAT$_{(SS)(12)}$} \\
\begin{tabular}{c||c|c|c|c|c|c||c}
\hline
& \textbf{15dB} & \textbf{10dB} &  \textbf{5dB} & \textbf{0dB} & \textbf{-5dB} &  \textbf{-10dB} & \textbf{Avg.}\\
\hline
CAR & \textbf{0.94} & \textbf{0.93} & \textbf{0.91} & 0.87 & \textbf{0.83} & \textbf{0.77} & \textbf{0.88}\\
\hline
PINK & 0.94 & \textbf{0.93} & \textbf{0.90} & \textbf{0.85} & \textbf{0.75} & \textbf{0.60} & \textbf{0.83}\\
\hline
STREET & 0.94 & \textbf{0.93} & \textbf{0.90} & \textbf{0.85} & \textbf{0.76} & \textbf{0.64} & \textbf{0.84}\\
\hline
BABBLE & 0.94 & \textbf{0.93} & 0.89 & \textbf{0.82} & \textbf{0.69} & \textbf{0.51} & \textbf{0.80}\\
\hline
\hline
\textbf{Avg.}  & \textbf{0.94} & \textbf{0.93} & \textbf{0.90} & \textbf{0.85} & \textbf{0.76} & \textbf{0.63} & \textbf{0.83}\\
\end{tabular} \\
\hrulefill \\
\hrulefill \\
\textbf{DAEME-USAT$_{(WD)(12)}$} \\
\begin{tabular}{c||c|c|c|c|c|c||c}
\hline
& \textbf{15dB} & \textbf{10dB} &  \textbf{5dB} & \textbf{0dB} & \textbf{-5dB} &  \textbf{-10dB}  & \textbf{Avg.}\\
\hline
CAR & \textbf{0.94} & \textbf{0.93} & \textbf{0.91} & \textbf{0.88} & \textbf{0.83} & \textbf{0.77}  & \textbf{0.88}\\
\hline
PINK & \textbf{0.95} & \textbf{0.93} & \textbf{0.90} & 0.84 & 0.74 & 0.59 & \textbf{0.83}\\
\hline
STREET & \textbf{0.95} & \textbf{0.93} & \textbf{0.90} & \textbf{0.85} & 0.75 & 0.62 & \textbf{0.83}\\
\hline
BABBLE & \textbf{0.95} & \textbf{0.93} & \textbf{0.90} & \textbf{0.82} & 0.68 & 0.50 & \textbf{0.80}\\
\hline
\hline
\textbf{Avg.}  & \textbf{0.95} & \textbf{0.93} & \textbf{0.90} & \textbf{0.85} & \textbf{0.75} & \textbf{0.62} & \textbf{0.83}\\
\end{tabular} \\
\hrulefill
\end{center}
\label{tab:STOI_DAEME-UAT12} 
\end{table}
\par

\subsection{Experiments on the TMHINT dataset}
The TMHINT corpus consists of speech utterances of eight speakers (four male and four female), each utterance corresponding to a sentence of ten Chinese characters. The speech utterances were recorded in a recording studio at a sampling rate of 16 kHz. Among the recorded utterances, 1,200 utterances pronounced by three male and three female speakers (each providing 200 utterances) were used for training. 120 utterances pronounced by another two speakers (one male and one female) were used for testing. There is no overlap between the training and testing speakers and speech contents. We used 100 different noise types \cite{100noise} to prepare the noisy training data at 31 SNR levels (from -10dB to 20 dB, with a step of 1 dB). Each clean utterance was contaminated by several randomly selected noise conditions (one condition corresponds to a specific noise type and an  SNR level). Finally, we collected 120,000 noisy-clean utterance pairs for training. For the testing data, four types of noises, including two stationary types (i.e., car and pink) and two non-stationary types (i.e., street and babble), were used to artificially generate noisy speech utterances at six SNR levels (-10 dB, -5dB, 0 dB, 5 dB, 10 dB, and 15 dB). As with the setup for the WSJ task, these four noise types were not included in preparing the training set. 

\graphicspath{ {./images/} }
\begin{figure}[t]
\centering
   {
\transparent{0.8}\includegraphics[width=8.8cm]{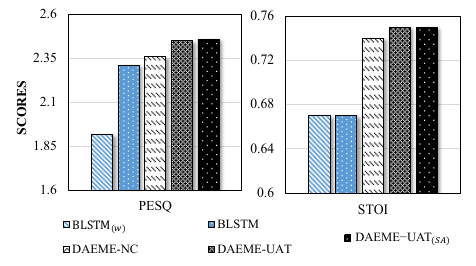} 
}
    \caption{Average PESQ and STOI scores of BLSTM$_{(w)}$, BLSTM, DAEME-NC, DAEME-UAT, and DAEME-UAT$_{(SA)}$.}
    \label{fig:knowledgecompare}
\end{figure}

\par
As we have evaluated the DAEME systems in several different aspects in the previous subsection, we will examine the DAEME algorithm in other aspects (including the achievable performance by incorporating other knowledge, for seen noise types, different component model types, different decoder model types,  and the ASR and listening test results) in this subsection. 

\subsubsection{Incorporating noise type and speaker identity information}
First, we implemented the DAEME by incorporating other knowledge, namely the noise type and speaker identity. In previous experiments, we assume that the gender and SNR information is available for each training utterance, and the UAT tree can be built based on such information. To compare, we built a DAEME system that prepares the component models without using the gender and SNR information. In this comparative system, we first built an utterance-based noise-type classifier by using a 2-layered BLSTM followed by one fully-connected layer. The input and output to the noise-type classifier, respectively, were a sequence of 257-dimensional LPS vectors and a 100-dimensional one-hot-vector (corresponding to the 100 noise types in the training set). Given an utterance, the noise-type classifier predicted the noise type of the utterance. We fed the entire set of training utterances into this noise-type classifier and collected the latent representations (each with 200 dimensions) of all of the training utterances. The entire set of latent representations were then clustered into $J$ clusters. Each cluster represented a particular cluster of training utterances and was used to train a component model. Finally, a fusion network was trained. Because a noise-classifier is used to partition the training data, this model is termed DAEME-NC. Different from the DAEME systems presented earlier, DAEME-NC prepared the component models in a data-driven manner without using the gender and SNR information. Because we intended to compare this approach with the UAT, we used 6 component models here for a fair comparison. The results of DAEME-NC are shown in Fig. \ref{fig:knowledgecompare}. The results of BLSTM and DAEME-UAT are also listed for comparison. From the figure, we can note that DAEME with knowledge-based UAT outperforms DAEME-NC, while DAEME-NC provides better performance than a single BLSTM model, in terms of both STOI and PESQ. All the improvements have been confirmed significant with very small $p$-values.

In \cite{chuang2019speaker}, a speaker-aware SE system was proposed and shown to provide improved performance. In this work, we conducted an additional set of experiments that integrate the speaker-aware techniques into the DAEME (termed DAEME-UAT$_{(SA)}$). For DAEME-UAT$_{(SA)}$, we included the speaker information in the decoder used on the same approach proposed in \cite{chuang2019speaker}. More specifically, we first extracted embedded speaker identity features using a pre-trained DNN model, which was trained to classify a frame-wise speech feature into a certain speaker identity. Then, the decoder of DAEME took augmented inputs (a combination of the outputs from the multi-branched encoder and the identity features) to generate enhanced spectral features. With the additional identity features, the overall enhancement system can be guided to generate the outputs considering the speaker identity. The results of DAEME-UAT$_{(SA)}$ are also listed in Fig. \ref{fig:knowledgecompare}. DAEME-UAT$_{(SA)}$ yields better performance than DAEME-UAT. The improvements are significant with very small $p$-values. The result confirms the effectiveness of incorporating the speaker information into the DAEME system.

We further investigated the advantage of the ensemble learning used in DAEME and thus trained a very wide and complex BLSTM model, which has exactly the same number of parameters as DAEME-UAT. The wide and complex BLSTM is termed BLSTM$_{(W)}$, and the results are also reported in Fig. \ref{fig:knowledgecompare}. From the figure, we note that the performance of BLSTM$_{(W)}$ is much worse than DAEME-UAT and other systems. The results well present a major advantage of the ensemble learning concept: to implement the target task, it is not needed to train a very wide, big, and complex model, which may take a lot of computation power and resource, yet difficult to train. Instead, we can train a set of simpler (weak) component models and a fusion network, which combines the outputs of these simple component models. Moreover, in terms of incorporating new knowledge, the ensemble learning approach has much more flexibility than a universal and complex model.

\subsubsection{Evaluation under seen noise types}
Next, we examine the DAEME effectiveness under seen noise conditions. We prepared testing data that involved two noise types (Cafeteria and Crowd) that were included in the training set. Then, we tested the enhancement performance of DAEME on the above testing noisy data. The PESQ and STOI results are listed in Tables \ref{tab:PESQ_DAEME_seen_data} and \ref{tab:STOI_DAEME_seen_data}, respectively. For comparison, we also listed the scores of BLSTM. The results in the tables show that DAEME outperforms the single BLSTM model with a clear margin. The improvements have been confirmed significant with very small $p$-values. The result confirms that DAEME, which preserves local information of the training data, can yield better performance than BLSTM, in which the local information may be averaged out, for seen noise types.
\begin{table}[htb]
\caption{\scshape{P}ESQ scores of seen data of BLSTM and DAEME.}
\begin{center} 
\begin{tabular}{c||c|c|c}
\hline
& \textbf{Noisy} & \textbf{BLSTM} &  \textbf{DAEME} \\
\hline
CAFETERIA & 2.01 & 2.56 & \textbf{2.70} \\
\hline
CROWD & 1.99 & 2.55 & \textbf{2.70} \\
\hline
\hline
\textbf{Avg.} & 2.00 & 2.56 & \textbf{2.70} \\
\hline
\end{tabular} \\
\end{center}
\label{tab:PESQ_DAEME_seen_data}

\caption{\scshape{S}TOI scores of seen data of BLSTM and DAEME.}
\begin{center} 
\begin{tabular}{c||c|c|c}
\hline
& \textbf{Noisy} & \textbf{BLSTM} &  \textbf{DAEME} \\
\hline
CAFETERIA & 0.65 & 0.71 & \textbf{0.72} \\
\hline
CROWD & 0.66 & 0.72 & \textbf{0.74} \\
\hline
\hline
\textbf{Avg.} & 0.65 & 0.71 & \textbf{0.73} \\
\hline
\end{tabular} \\
\end{center}
\label{tab:STOI_DAEME_seen_data} 
\end{table}
\par


\graphicspath{ {./images/} }
\begin{figure}[t]
\centering
   \includegraphics[width=8.8cm]{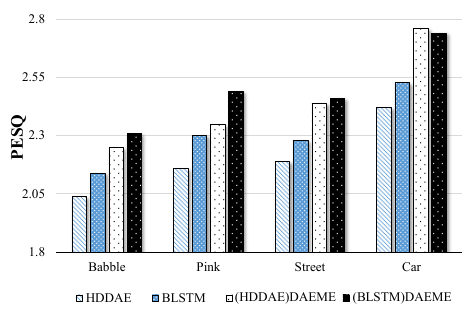}
    \caption{PESQ scores of SE systems using different component models in the multi-branched encoder. The scores of unprocessed noisy speech are 1.94, 1.78, 1.95, and 2.36 for Babble, Pink, Street, and Car, respectively.}
\vspace{0.5cm}
\centering
   \includegraphics[width=8.8cm]{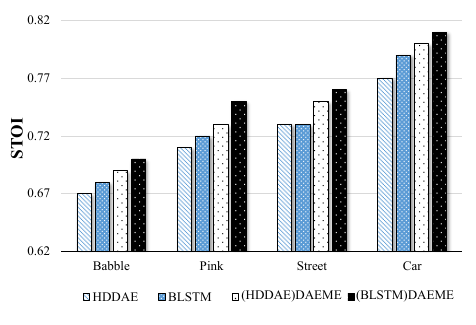}
    \caption{STOI scores of SE systems using different component models in the multi-branched encoder. The scores of unprocessed noisy speech are 0.69, 0.71, 0.72, and 0.79 for Babble, Pink, Street, and Car, respectively.}
\end{figure}

\graphicspath{ {./images/} }
\begin{figure}[t]
\centering
   \includegraphics[width=8.8cm]{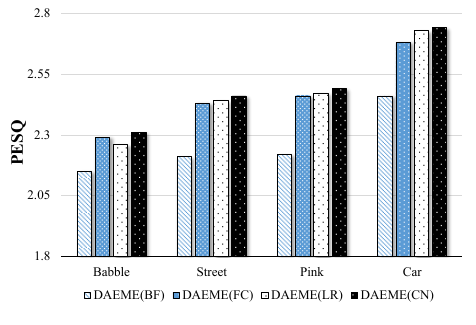}
    \caption{PESQ scores of DAEME-based ensemble SE systems using different types of decoder.}

\centering
   \includegraphics[width=8.8cm]{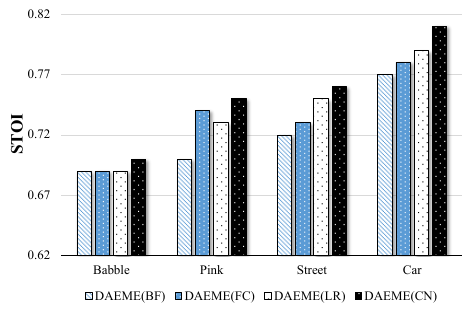}
    \caption{STOI scores of DAEME-based ensemble SE systems using different types of decoder.}
\end{figure}
\subsubsection{The component models of DAEME}
We analyzed the compatibility of the DAEME algorithm with different component models. In the previous experiments, we used BLSTM to build the multi-branched encoder and CNN as the decoder. First, we followed the setup of the best model DAEME-USAT$_{(WD)(12)}$ and adopted HDDAE in addition to BLSTM as the architecture of a component model. The BLSTM-based component model consisted of two layers, with 300 memory cells in each layer, and the HDDAE-based component model consisted of five hidden layers, each containing 2,048 neurons, and one highway layer \cite{ensemble5}. Figs. 9 and 10 present the PESQ and STOI scores of different systems, including two single model SE systems (HDDAE, and BLSTM) and two DAEME-based ensemble systems, (BLSTM)DAEME, and (HDDAE)DAEME, where BLSTM, and HDDAE were used to build the component models in the multi-branched encoder, respectively. From the figures, we can note that (HDDAE)DAEME, and (BLSTM)DAEME outperform their single model counterparts (HDDAE and BLSTM). It is also shown that among the two single model-based systems, BLSTM outperformed HDDAE. Among the two DAEME-based systems, (BLSTM)DAEME achieves the best performance, suggesting that the architecture of the component model of DAEME indeed affects the overall performance.  
\par
\subsubsection{The architecture of decoder}
Next, we investigated the decoder in the DAEME-based systems. In this set of experiments, BLSTM with the same architecture as in the previous experiments was used to build the component SE models. We compared four types of decoder: the CNN model used in the previous experiments and the linear regression as shown in Eq.(1). We used another nonlinear mapping function based on a fully connected network. Finally, we used the BF approach, which selects the most suitable output from the component models as the final output. More specifically, we assume that the gender and SNR information were accessible for each testing utterance, and the BF approach simply selected the "matched" output without further processing. The results of using different decoder types are listed in Figs. 11 and 12, which report the PESQ and STOI scores, respectively. For DAEME(LR), the decoder is formed by a linear regression function, for DAEME(FC) and DAEME(CN), the decoder is formed by a nonlinear mapping function based on a fully connected network and CNN, respectively, and for DAEME(BF), the decoder is formed by the BF approach. Please note that the way to combine multiple outputs of encoders depends on the type of decoder. When using BF as the decoder, only one output from the multi-branched encoder is used. When using LR, FC, or CN as the decoder, we first concatenate the encoded feature maps, and then compute the enhanced speech. We followed Eq. (1) to implement LR. For FC, we used 2 dense layers, each with 1024 nodes. For CNN, we used 3 layers of 1-D convolution, each with kernel size 11, stride 1, and 64 channels, followed by 2 dense layers, each with 1024 nodes. No pooling layer was used for CNN.

From Figs. 11 and 12, it is clear that DAEME(CN) outperforms DAEME(FC), DAEME(LR), and DAEME(BF) consistently. The results confirm that when using a decoder that has more powerful nonlinear mapping capability, the DAEME can achieve better performance. We also conducted the dependent t-test to verify whether the improvements are significance. It is confirmed that the improvements of DAEME(CN) over BLSTM are significant in terms of both PESQ and STOI scores with very small $p$-values, again verifying the effectiveness of the DAEME model for providing better speech quality and intelligibility over a single model.

\subsubsection{ASR performance}
\begin{figure}[t]
\centering
   \includegraphics[width=8cm]{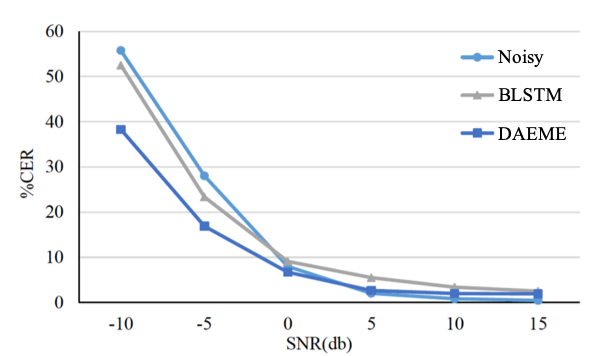}  
    \caption{CERs of Google ASR applied to noisy and enhanced speech.}
    \label{fig:CER}
\end{figure}

The previous experiments have demonstrated the ability of the DAEME methods to enhance the PESQ and STOI scores. Here, we evaluated the applicability of DAEME as a denoising front end for ASR under noisy conditions. Google ASR \cite{ASR, donahue2018exploring} was adopted to test the character error rate (CER) with the correct transcription reference. The best setup of DAEME for the TMHINT dataset, i.e., DAEME-USAT$_{(WD)(12)}$, was used to pre-process the input noisy speech, and the enhanced speech was sent to Google ASR. The unprocessed noisy speech and the enhanced speech by a single BLSTM model were also tested for comparison. The CER results for these three experimental setups are demonstrated in Fig. \ref{fig:CER}. It is clear that the single-model BLSTM SE system is not always helpful. The enhanced speech tends to yield higher CERs than the unprocessed noisy speech while tested under higher SNR conditions (0 to 15dB) and achieve only slightly lower CERs under relatively noisier conditions (-10 to 0dB). On the other hand, the proposed DAEME system achieves lower CERs under lower SNR conditions (-10 to 0dB) and maintains the recognition performance under higher SNR conditions (0 to 15dB) as compared to the noisy speech. The average CERs of noisy, BLSTM-enhanced, and DAEME-enhanced speech across 15dB to -10dB SNR levels are 15.85\%, 16.05\%, and 11.4\%, accordingly. DAEME achieved a relative CER reduction of 28.07\% (from 15.85\% to 11.4\%) over the unprocessed noisy speech. 

\graphicspath{ {./images/} }

\subsubsection{Listening test}
In addition to the objective evaluations and ASR tests, we also invited 15 volunteers to conduct subjective listening tests. The testing conditions included two types of noises, car noise (stationary noise) and babble noise (non-stationary noise), under two different SNR levels (car: -5dB, -10dB, babble: 0dB, 5dB). We selected different SNRs for these two noise types because the car noise is more stationary and thus easier understood as compared to the babble noise; accordingly a lower SNR should be specified for the car noise than the babble noise.\par
\begin{figure}[t]
 \centering
 \subfigure[SIG scores.]{
   \includegraphics[height = 2.4cm, width = 2.62cm]{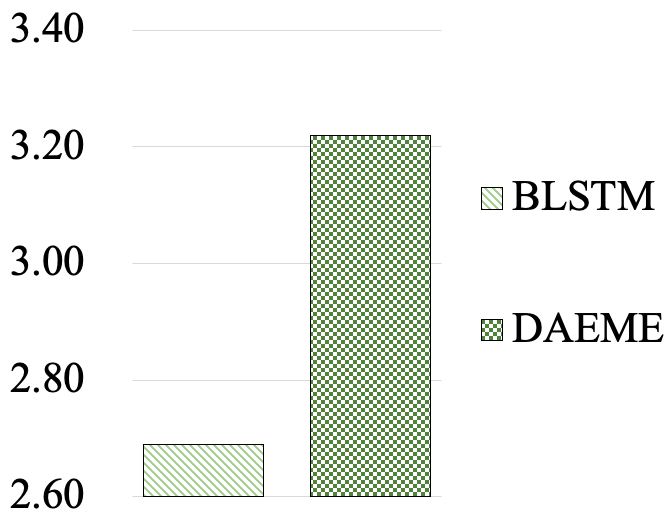}
   \label{fig:sigscore}
 }
 \subfigure[BAK scores.]{
   \includegraphics[height = 2.4cm, width = 2.62cm]{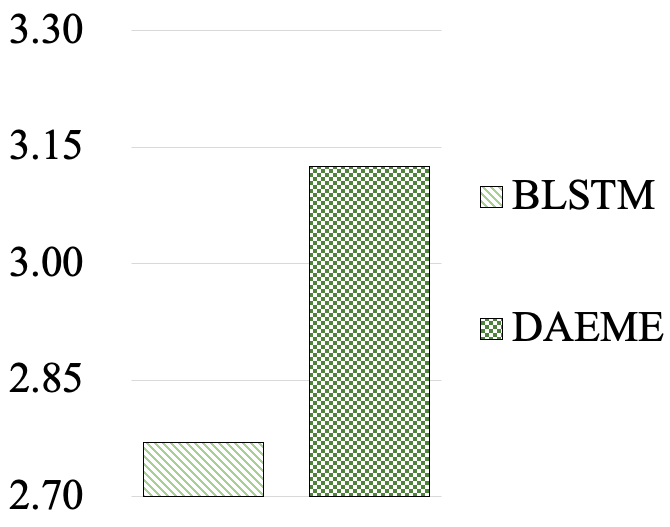}
   \label{fig:bakscore}
 }
  \subfigure[Preference scores.]{
   \includegraphics[height = 2.4cm, width = 2.62cm]{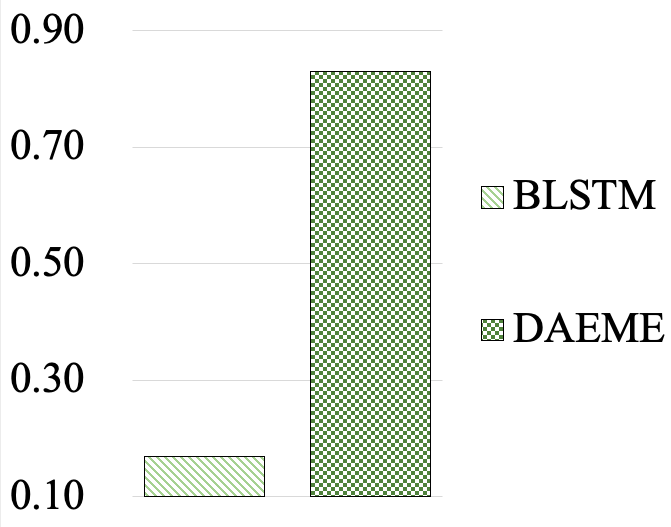}
   \label{fig:prefscore}
 }
\caption{Listening test results in terms of SIG, BAK, and AB preference.}
\end{figure}

 
First, we asked each subject to register their judgment based on the five-point scale of signal distortion (SIG) \cite{BAK}. In the SIG test, each subject was asked to provide the natural (no degradation) level after listening to an enhanced speech utterance processed by either BLSTM or DAEME. A higher score indicates that the speech signals are more natural. The results shown in Fig. \ref{fig:sigscore} demonstrate that DAEME can yields a higher SIG score than BLSTM. We also asked the subjects to judge the background intrusiveness (BAK) \cite{BAK} after listening to an utterance. The BAK score ranges from 1 to 5, and a higher BAK score indicates a lower level of noise artifact perceived. The BLSTM and DAEME enhanced utterances were tested for comparison. The results shown in Fig. \ref{fig:bakscore} clearly demonstrate that DAEME outperforms BLSTM in terms of BAK. Finally, we conducted an AB preference test \cite{Loizou:2013:SET:2484638} to compare the BLSTM-enhanced speech and the DAEME-enhanced speech. Each subject listened to a pair of enhanced speech utterances and gave a preference score. As shown in Fig. \ref{fig:prefscore}, the DAEME notably outperforms BLSTM in the preference test. The results in Fig. \ref{fig:sigscore}, \ref{fig:bakscore}, and \ref{fig:prefscore} demonstrate that, compared to BLSTM, the speech enhanced by DAEME is more natural, less noisy, and with higher quality.        
 \graphicspath{ {./images/} }

\section{Conclusions}
\par
This paper proposed a novel DAEME SE approach. By considering the ensemble concept, the proposed method first exploits the complimentary information based on the multi-branched encoder, and then uses a decoder to combine the complementary information for SE. We also analyzed and confirmed that the decoder using CNN-based non-linear transformation yielded better SE performance than the decoder using 2-layered fully connected network, linear transformation and the BF approach. Compared to other learning-based SE approaches, the proposed DAEME approach has the following three advantages: 
\subsubsection{\textcolor{black}{The DAEME approach yields a better enhancement performance under both seen and unseen noise types than its baseline single model counterparts}}
As presented in Section III-A, the DSDT was built based on the utterance-level attributes (UA) and signal-level attributes (SA) of speech signals. The DSDT is subsequently used to establish the multi-branched encoder in the DAEME framework. From the PESQ and STOI scores, we confirmed the effectiveness of the UA tree and the USAT (UA+SA) tree over traditional deep-learning-based approaches and the system using a random tree (RT in Figs. 6 and 7). The experimental results also confirm that the proposed DAEME has effectively incorporated the prior knowledge of speech signals to attain improved enhancement performance as compared to single model counterparts under both seen and unseen noise types.  
\subsubsection{The DAEME architecture with a DSDT has better interpretability and can be designed according to the amount of training data and computation resource constraints}
Pre-analyzing the speech attributes of the target domain using a regression tree provides informative insights into the development of the DAEME system. By using the tree map that categorized data through speech attributes, we can optimally design the architecture of the multi-branched encoder in the DAEME to strike a good balance between the achievable performance, training data, and the system complexity. \textcolor{black}{The component models can be trained in parallel using respective training subsets. Because multiple component models can handle different noise conditions well, a simple decoder that is easy to train can already provide satisfactory performance. This is confirmed by the experimental results Figs. 11 and 12. Therefore, the training of the fusion layer (decoder) will not require much training time. Compared to BLSTM, CNN can be much simplified in terms of hardware and computation costs. Therefore, BLSTM plays a major part in the total complexity of DAEME. According to \cite{tsironi2017analysis}, the time complexity of BLSTM can be derived as $O((CH)^2)$, where $H$ is the basic number of cells and $C$ is the width factor. We can arrange the computation of component models in a parallel manner that further lowers complexity. For example, the time complexity of a 6x wider BLSTM model is high ($C$=6), while that of DAEME can be much reduced by a parallel computation.}
\subsubsection{The DAEME system can incorporate noise-type and speaker identity information to further improve the performance} As presented in Section IV-D-1, we can build the component models based on the clustering results of latent representations of the training utterances. In this way, data clustering is implemented in a data-driven manner without the need of gender and SNR level information. Moreover, we have tried to incorporate the speaker identity information into DAEME (cf. Section IV-D-1)). The results also show that the SE capability can be further improved by incorporating the speaker identity information. As compared to a big and complex universal model, DAEME has more flexibility to incorporate new data-driven or acoustic information.

In the future, we will explore to apply the proposed DAEME to other speech signal processing tasks, such as speech dereverberation and audio separation. Meanwhile, in the present study, we did not consider to automatically determine a suitable number of models based on the testing scenarios. \textcolor{black}{That is, the number of component models must be specified in the training stage. An algorithm that can online determine the optimal encoder architecture based on the complexity and performance also deserves further investigations.} Finally, we will implement the proposed DAEME system on practical devices for real-world speech-related applications.  





%









\bibliographystyle{IEEEbib}

\bibliography{refs}
\section{Appendix}
We summarize the abbreviations used in this article:
\textcolor {black} {
\begin{itemize}
\item DAEME: denoising autoencoder with multi-branched encoders.
\item HDDAE: DDAE with a highway structure.
\item UAT: utterance-level attribute.
\item SAT: signal-level attribute.
\item USAT: utterance- and signal-level attribute.
\item DSDT: dynamically-sized decision tree.
\item RT:  random tree.
\item SS: spectral segmentation. 
\item WD: wavelet decomposition.
\item BF: the best first based decoder in DAEME.
\item FC: the fully-connected network based decoder in DAEME.
\item LR: the linear regression based decoder in DAEME.
\item CN: the CNN based decoder in DAEME.
\end{itemize}
}

\end{document}